\documentclass{PoS}
\def\Journal#1#2#3#4{{#1} {\bf #2}, #3 (#4)}

\def\PRL{\em Phys. Rev. Lett.}

\def\ra{\rightarrow}

\def\be{\begin{equation}}
\def\ee{\end{equation}}
\def\bea{\begin{eqnarray}}
\def\eea{\end{eqnarray}}

\title{Measurement of $\beta_s$ at CDF}

\ShortTitle{Measurement of $\beta_s$ at CDF}

\author{\speaker{Louise Oakes}\thanks{For the CDF collaboration}\\
        Oxford University\\
        E-mail: \email{loakes@fnal.gov}}

\abstract{The latest results for the measurement of the $CP$ violating phase $\beta_{s}$ in $B_s^0 \rightarrow J/\psi\phi$ decays, from 5.2 fb$^{-1}$ integrated luminosity of CDF data are presented.  For the first time, this measurement includes the contribution of $B_s^0 \rightarrow J/\psi K^+K^-$ or $B_s^0 \rightarrow J/\psi f_0$ events to the signal sample, where the $f_0$ and  non-resonant $K^+K^-$ are $S$-wave states. Additional improvements to the analysis include more than doubling the signal sample, improved selection and particle ID, and fully calibrated flavour tagging for the full dataset. Additionally, the world's most precise single measurements of the $B_{s}^{0}$ lifetime, $\tau_{s}$, and width difference, $\Delta\Gamma_{s}$ are given.  }

\FullConference{Flavor Physics and CP Violation - FPCP 2010\\
		May 25-29, 2010\\
		Turin, Italy}

\begin{document}

\section{Introduction}
The study of neutral $B$ meson properties can provide important tests of the Standard Model (SM) including constraints on parameters of the Cabibbo-Kobayashi-Maskawa (CKM) matrix.  While the $B^{0}$ system has been thoroughly investigated by $B$ factories, precision measurements in the $B^{0}_{s}$ system are a more recent development, driven largely by the Tevatron experiments.  The $B_{s}^{0}-\bar B_{s}^{0}$ system has the potential to yield indirect observations of New Physics (NP), through the presence of non-SM particles in second order weak interaction processes such as flavour mixing.  
The golden mode for this measurement is $B_{s}^{0}\rightarrow J/\psi\phi$. The $J/\psi\phi$ final state is common to $B_{s}^{0}$ and $\bar B_{s}^{0}$ decays; $CP$ violation occurs in the mixing through interference between decays with and without $B_{s}^{0}$ mixing.  The phase, $\beta_{s}$, between these two amplitudes is predicted to be close to zero in the SM~\cite{UTFit}, so a significant excess would be a clear indication of evidence for NP in this channel. 

The updated measurement is of particular interest as a published CDF $B_{s} \ra J/\psi\phi$ analysis and a recent update showed deviations from the Standard Model expected value of approximately 2 $\sigma$~\cite{CDFprd, CDFpub}, and a similar effect was seen in a recent combined Tevatron result~\cite{comb}.

\section{Neutral $B^{0}_{s}$ system phenomenology}
The flavour eigenstates of $B^{0}_{s}$ mesons in the SM are not the same as the mass eigenstates, leading to oscillations between $|B_{s}^{0} \rangle = (\bar b s) $ and $|\bar B_{s}^{0} \rangle = (b \bar s)$ via the second order weak interactions.  The phenomenology of this weak mixing is described by the CKM  matrix.  The time evolution of the $B^{0}_{s}-\bar B_{s}^{0}$ system is governed by the Schr\"odinger equation
\be
i\frac{d}{dt}\left(\begin{array}{c} B_{s}^{0}(t) \\ \bar   B_{s}^{0}(t) \end{array} \right ) 
= \mathcal{H}\left(\begin{array}{c} B_{s}^{0}(t) \\ \bar   B_{s}^{0}(t) \end{array} \right )   
\equiv \left [ 
\left(
\begin{array}{cc}  
M_{0} & M_{12} \\ 
M^{*}_{12} & M_{0}
\end{array}
\right) - \frac{i}{2} 
\left(
\begin{array}{cc}  
\Gamma_{0} & \Gamma_{12} \\ 
\Gamma^{*}_{12} & \Gamma_{0}
\end{array}
\right)
\right ]
\left(\begin{array}{c} B_{s}^{0}(t) \\ \bar   B_{s}^{0}(t) \end{array} \right )
\ee
where the $M$ and $\Gamma$ matrices describe the mass and decays of the system. The mass eigenstates can be obtained by diagonalising $\mathcal{H}$, giving:
\be 
|B_{s}^{H}\rangle = p|B_{s}^{0}\rangle - q|\bar B_{s}^{0}\rangle
\ee
and 
\be
|B_{s}^{L}\rangle = p|B_{s}^{0}\rangle + q|\bar B_{s}^{0}\rangle
\ee
where $|q/p|= 1$ in the case of no $CP$ violation in mixing (as opposed to between decays with and without mixing), as predicted in the $J/\psi \phi$  channel. The indices $H$ and $L$ label the heavy and light eigenstates respectively.  The mass difference, $\Delta m_{s}$, between the heavy and light states is proportional to the frequency of $B^{0}_{s}$ mixing and is approximately equal to $2|M_{12}|$. The mass eigenstates have a small but non negligible lifetime difference, which can be described in terms of the decay width difference 
$\Delta\Gamma_{s} = \Gamma_{L}-\Gamma_{H}  \approx 2|\Gamma_{12}|\cos(2\phi_{s}) $
where the $CP$ violating phase is defined as $\phi_{s} = arg(-M_{12}/\Gamma_{12})$ and the mean decay width $\Gamma_{s} =1/\tau_{s}$.  The SM predicts $\phi_{s}^{SM}$ to be of order 0.004, and it would be expected to increase in the presence of NP.  The mixing frequency, $\Delta m_{s}$ has been well determined by CDF, a $5\sigma$ observation of mixing with $\Delta m_{s}=17.77 \pm 0.10~ \textrm{(stat.)}\pm 0.07~\textrm{(syst.)}\rm{ps}^{-1}$ was published in 2006~\cite{CDFmix}. 

The full angular and time dependent equations for the measurement of $B_s^0 \rightarrow J/\psi\phi$, including the addition of the $S$-wave $KK$ component are detailed in~\cite{nim}.

\section{\textit{CP} violation in $B_{s}\rightarrow J/\psi\phi$}
The relative phase, $\beta_{s}$, between decays of a $B_{s}^{0}$ meson to $J/\psi \phi$ directly, and after mixing to $\bar B_{s}^{0}$, is defined in the SM as 
\be
\beta_{s}^{SM} = arg\left(\frac{-V_{ts}V_{tb}^{*}}{V_{cs}V_{cb}^{*}}\right) \approx 0.02
\ee
A New Physics phase, contributing to the weak mixing diagrams in the neutral $B_{s}^{0}$ system would introduce a new physics phase $\phi_{s}^{NP}$ to $\beta_{s}$ such that the measured value would be $2\beta_{s} = 2\beta_{s}^{SM} - \phi_{s}^{NP}$.  The same NP phase would enhance $\phi_{s}$, giving $\phi_{s} = \phi_{s}^{SM} + \phi_{s}^{NP}$.  As both $\beta_{s}^{SM}$ and $\phi_{s}^{SM}$ are predicted to be close to zero, the NP phase would dominate, and the measured phase would be~$2\beta_{s} \approx -\phi_{s} \approx \phi_{s}^{NP}$.  This approximation is valid given the current experimental resolution, but future high precision measurements may be able to distinguish between these quantities.

\subsection{Experimental strategy}
\begin{figure}[htb]
\begin{center}
        \includegraphics[width=8cm]{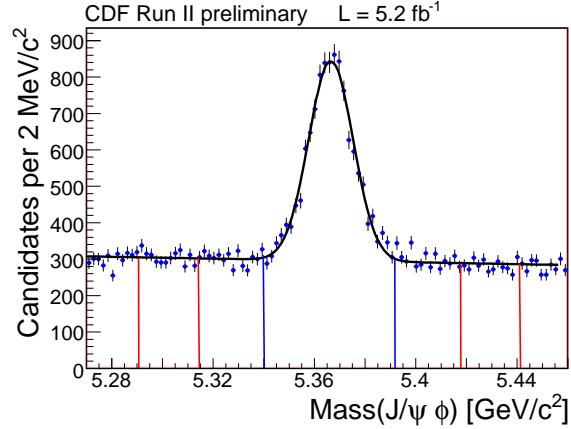}
\end{center}
\caption{$B_s^0$ mass distribution for CDF 5.2 fb $^-1$ data sample, fitted with a single Gaussian for the signal and a 1st order polynomial to describe the background component.}
\label{fig:yield}
\end{figure}

The decay $B_{s}^{0} \rightarrow J/\psi(\rightarrow \mu^{+} \mu^{-})\phi(\rightarrow K^{+}K^{-})$ is fully reconstructed from events which pass the di-muon trigger.  A Neural Network (NN) selection procedure is used to reconstruct $\sim$ 6500 signal events in 5.2 $^{-1}$ of data. For this updated analysis, rather than using a standard $S/\sqrt(S+B)$ figure of merit, the NN cut is optimised by selecting the cut value which minimises the statistical error on $\beta_{s}$ in pseudo experiments. This, along with a full re-calibration of particle identification information from $dE/dx$ and Time of Flight (TOF), has lead to a $B_s^0\rightarrow J/\psi\phi$ signal yield of more than twice that in the previous 2.8 fb$^{-1}$ data sample~\cite{CDFpub}. Figure~\ref{fig:yield} shows the $B_s^0\rightarrow J/\psi\phi$ mass distribution after selection.

%

The final state is an admixture of $CP$ odd and even states, which can be separated according to their angular momentum.  The total angular momentum of the $J/\psi \phi$ state can be $L=$ 0, 1, or 2, and the $CP$ of the state is $(-1)^{L}$, so the $L=0,2$ states are $CP$ even, and the $L=1$ state is $CP$ odd.  
These $CP$ states can be separated using the angular distribution of the four final state particles, the muons and kaons from the decay of the $J/\psi$ and $\phi$. The transversity basis~\cite{trans} is used to define the angular dependence of the final state, where the relative directions of the four particles can be described in terms of three transversity angles, $\{\cos\theta_{T}, \phi_{T}, \cos\psi_{T}\}$ which are defined by the direction of the decaying $J/\psi$ and $\phi$ mesons.  In the transversity basis, the decay amplitude can be separated into three components which represent different linear polarisation states.  The $CP$ even states correspond to the vector mesons either being longitudinally polarised, or transverse to their direction of motion and parallel to each other (0, $\parallel$), for the $CP$ odd state the mesons are transversely polarised with respect to their direction of motion, and perpendicular to each other ($\perp$). The amplitudes of these states are $A_{0}, A_{\parallel}$ and $A_{\perp}$ respectively.  The use of the transversity basis to separate the $CP$ odd and even states in this way means that the measurement is sensitive to the $CP$ violating phase with or without flavour tagging information for the initial state.

The angular analysis is combined with time development and mass dependence in a multivariate likelihood fit. In the simplest case, the fit without flavour tagging information, the likelihood function has a four fold ambiguity under the transformations $\{\beta_{s}, \Delta\Gamma, \phi_{\parallel}, \phi_{\perp}\} \Leftrightarrow \{\phi/2 - \beta_{s}, -\Delta\Gamma,2\pi- \phi_{\parallel},\pi- \phi_{\perp}\}$ and $\beta_{s} \Leftrightarrow -\beta_{s}$, where the strong phases are defined in terms of the transversity amplitudes, $\phi_{\parallel} \equiv arg(A^{*}_{\parallel}A_{0})$ and $\phi_{\perp} \equiv arg(A^{*}_{\perp}A_{0})$. 

\subsection{Flavour tagging}
By flavour tagging the initial $B_{s}^{0}$ meson, the time development of $B_{s}^{0}$ and $\bar B_{s}^{0}$ states can be followed separately, which removes the insensitivity to the sign of $\beta_{s}$ and $\Delta\Gamma$. This reduces the ambiguity to two points. The flavour of the decaying $B$ meson is tagged using a combination of opposite side (OST) and same side (SST) tagging algorithms.  The OST tags on the $b$ quark content of a $B$ meson from the same production vertex as the candidate $B_{s}^{0}$, the SST tags according to the $s$ quark content of a kaon produced with the candidate.

For this updated analysis, the SST has been re-calibrated for the full sample on data through a $B^0_{s}$ mixing measurement. This technique uses the fact that a measured mixing amplitude of $\approx$ 1 means that the tagger accurately assesses its performance, and an amplitude of $>1$ or $<1$ implies an under or over estimation of its power, respectively. 

This calibration uses the modes:
\begin{eqnarray*}
B_s^0 &\ra& D_s^-\pi^+, D_s^- \ra \phi \pi^-, \phi \ra K^+K^- \\
B_s^0 &\ra& D_s^-\pi^+, D_s^- \ra  K^*K^- , K^* \ra K^+\pi^- \\
B_s^0 &\ra& D_s^-\pi^+, D_s^- \ra \pi^-\pi^-\pi^+ \\
B_s^0 &\ra& D_s^-\pi^+\pi^+\pi^-, D_s^- \ra \phi \pi^-, \phi \ra K^+K^- 
\end{eqnarray*}
Of which the first channel accounts for about 50\% of the statistics.

\begin{figure}[htb]
\begin{center}
        \includegraphics[width=8cm]{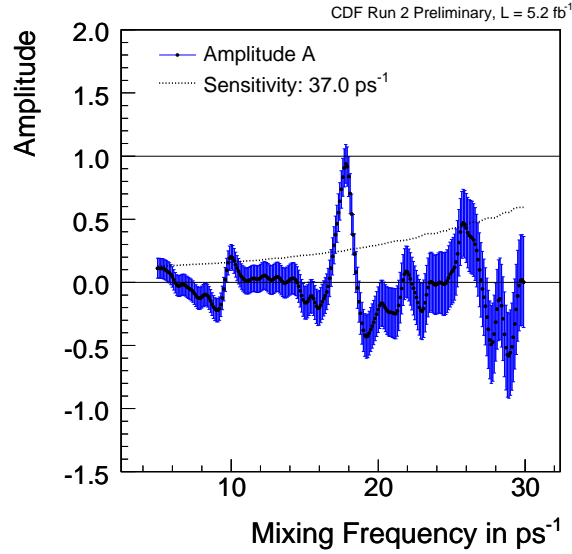}
\end{center}
        \caption{$B_{s}^{0}$ mixing amplitude scan,  for calibration of SST}
\label{fig:SSKT}
\end{figure}

The amplitude measured for this calibration~\cite{SSKT} is $\mathcal{A} = 0.94\pm0.15~\rm{(stat.)}\pm~0.13\rm{(syst.)}$, shown in Figure~\ref{fig:SSKT}. 
The mixing frequency, $\Delta m_{s} = 17.79 \pm 0.07 ~\rm{ps}^{-1}$, with statistical errors only, is in good agreement with the CDF published measurement.
%

\subsection{$S$-wave $KK$ component}
It has been suggested~\cite{stone} that a potential contamination of the signal $\phi$ meson by $S$-wave $f^{0}$ or non-resonant $KK$ of $\sim 10 \%$ could bias the measurement of $\beta_{s}$ towards the SM value. This updated CDF analysis includes the $S$-wave component in the full angular and time-dependent analysis.  Both the $f_0$ and non-resonant $KK$ components are considered flat in mass within the small selection window, and the $\phi$ meson mass is modelled by an asymmetric relativistic Breit Wigner; however this mass is integrated over in the fit function and is not used in the fit.  The $J/\psi K^+K^-$ or $J/\psi f_0$ final state is a pure $CP$ odd state, and thus follows the time dependence of the $CP$ odd component of the $B_s^0 \rightarrow J/\psi \phi$ decay.

\begin{figure}[htb]
\begin{center}
        \includegraphics[width=7cm]{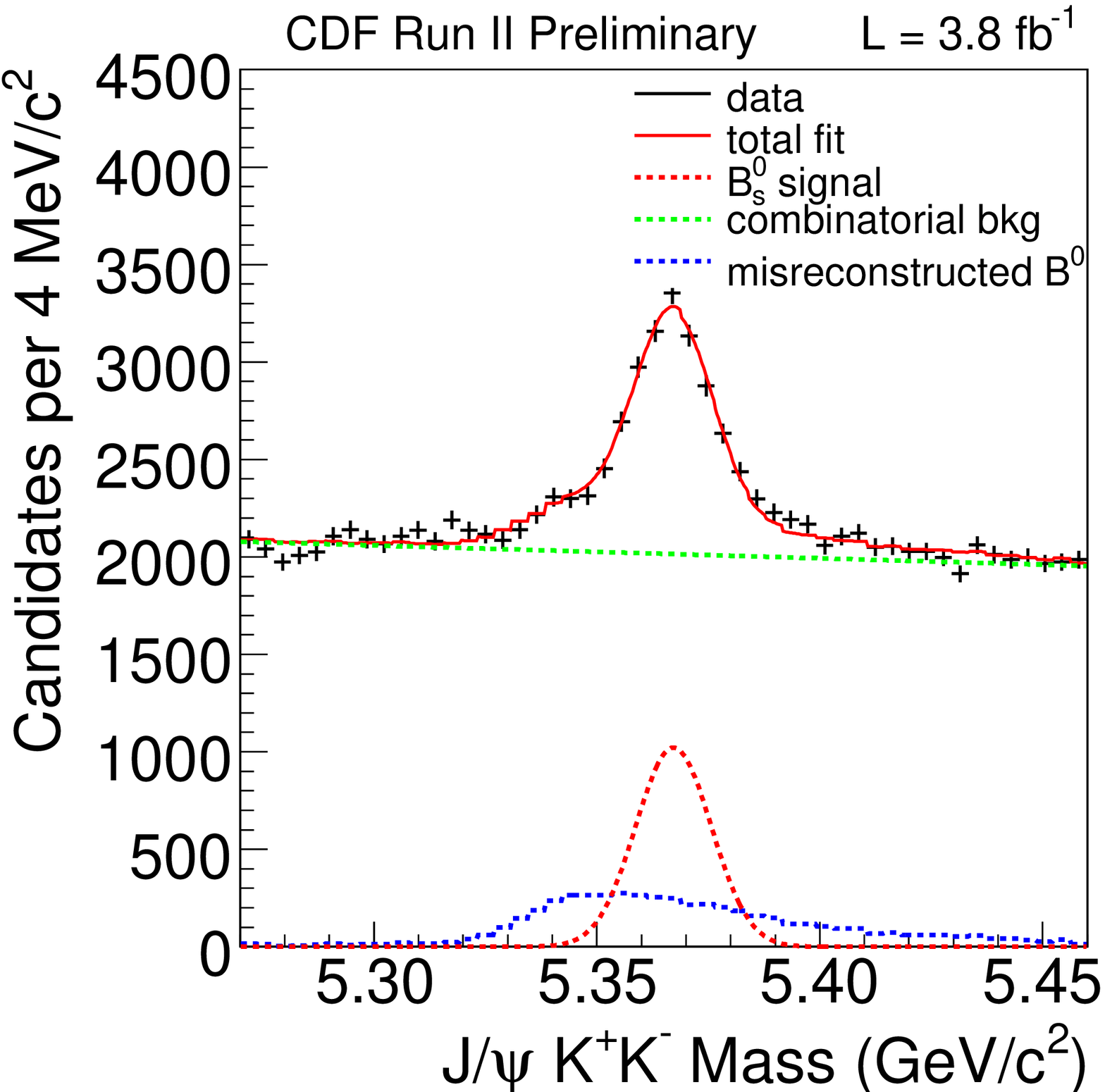}
 	\includegraphics[width=7cm]{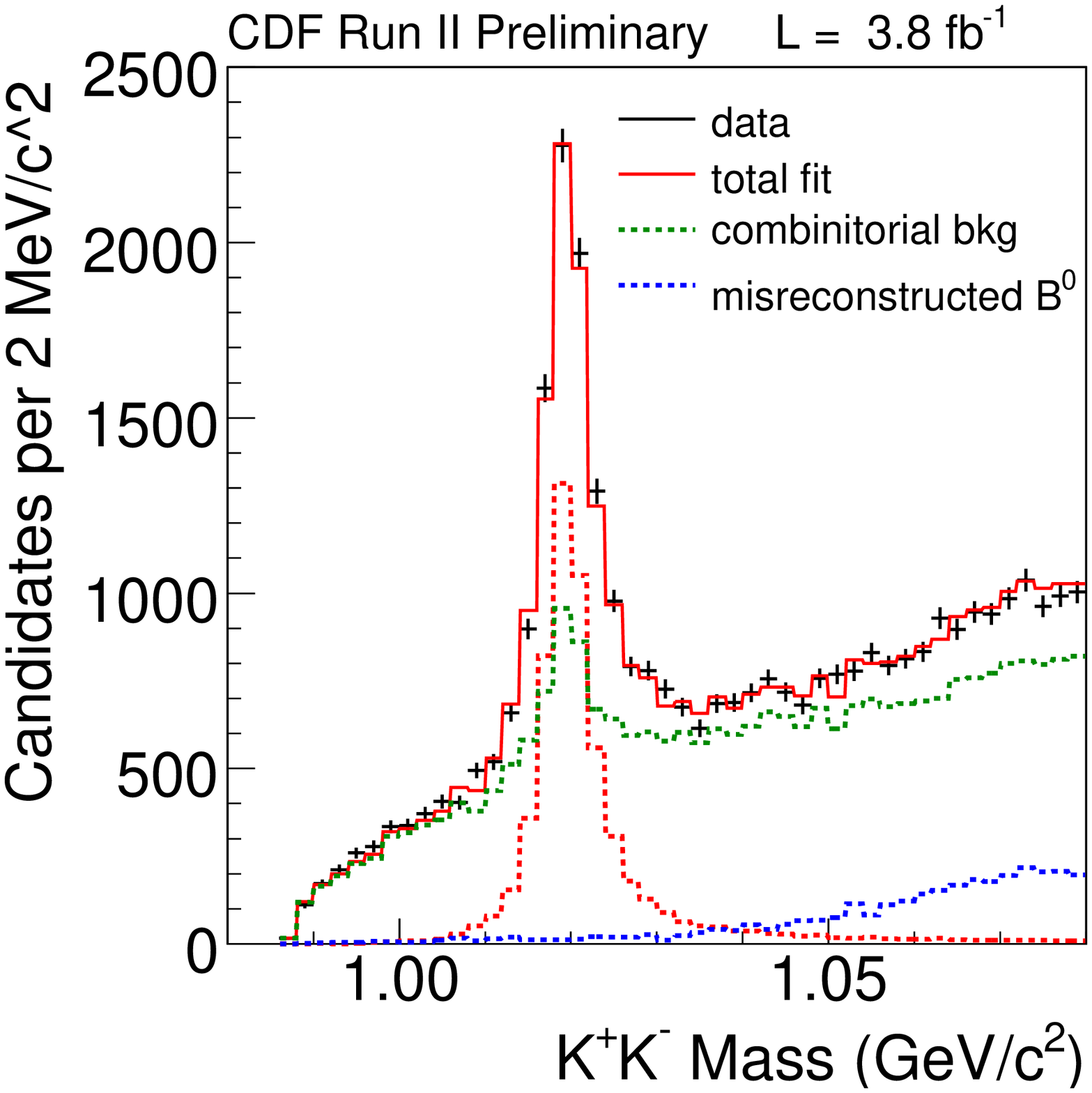}	
\end{center}
\caption{The $B^0_{s}$ mass is plotted (left) with a loose $\phi$ mass cut window, which allows contamination from $B^{0}\rightarrow J/\psi K^{*}$ misreconstructed as $B_{s}^{0} \rightarrow J/\psi \phi$, this reflection component is fitted with a MC template, the signal $B^0_{s}$ mass is fitted with a Gaussian and the combinatorial background with a 1st order polynomial. The invariant KK mass plot (right) is shown with the fractions of each component fixed to that found in the $B^0_{s}$ mass fit.}
\label{fig:updates}
\end{figure}

A preliminary study of the $S$-wave contamination of the $\phi$ meson signal carried out by studying the  invariant $KK$ mass distribution gives no strong indication of a large additional component, as shown in Figure~\ref{fig:updates}.
%

\section{Results}
\begin{figure}[htb]
\begin{center}
        \includegraphics[width=7cm]{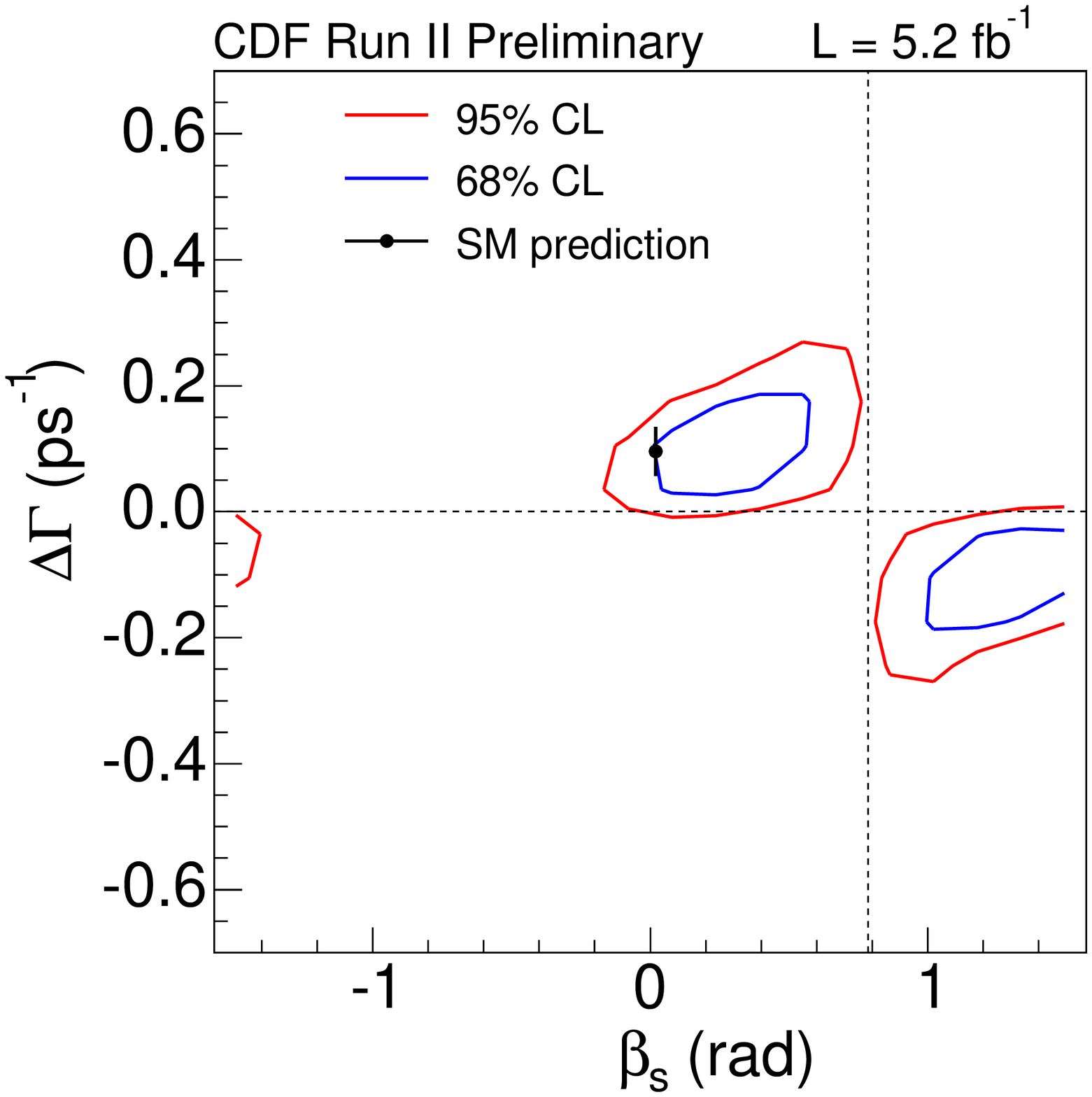}
 	\includegraphics[width=7cm]{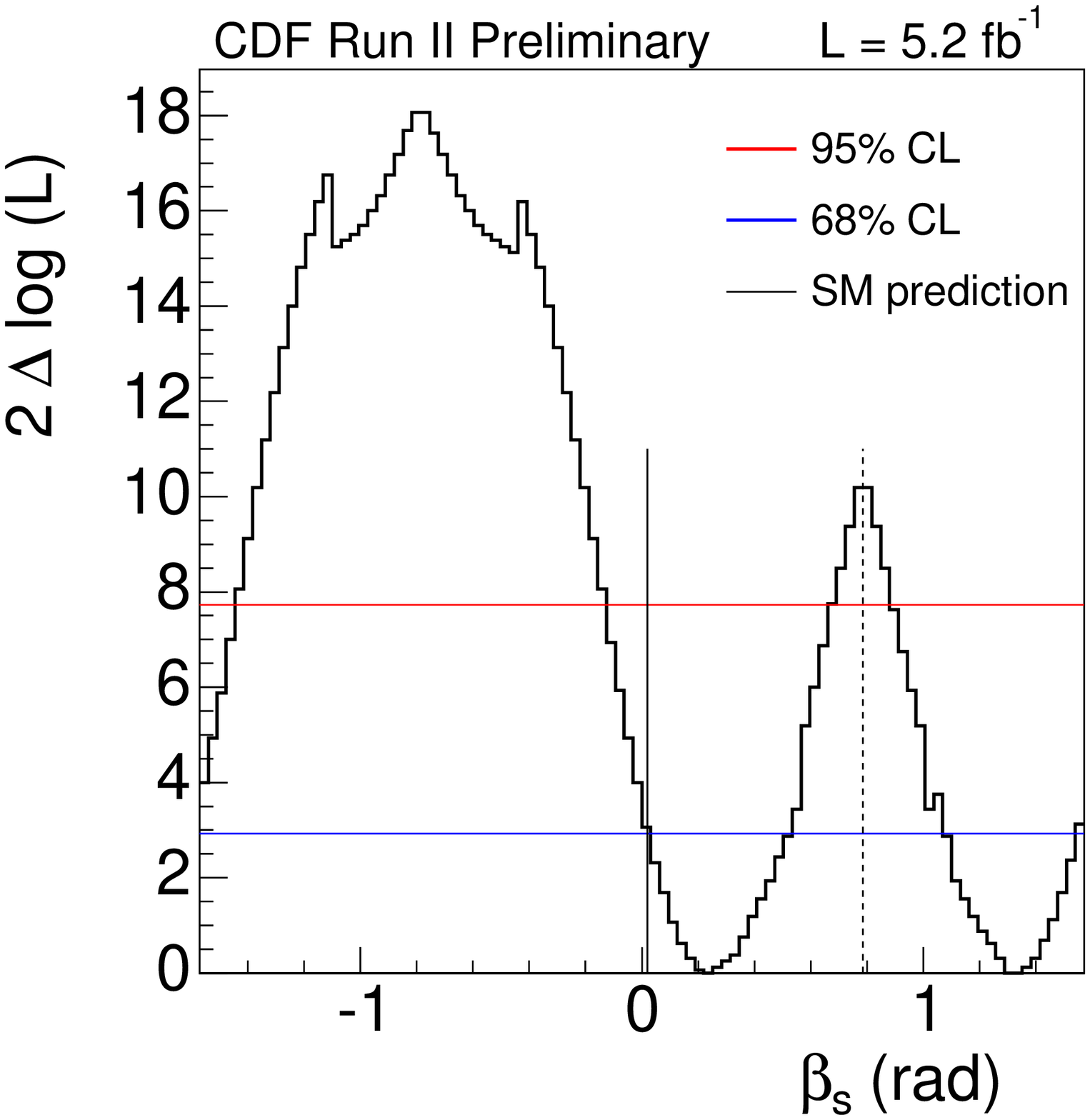}	
\end{center}
\caption{Confidence regions in 2-dimensions in the $\beta_{s}^{J/\psi\phi}-\Delta\Gamma_{s}$ plane, including systematics effects on the statistical uncertainties (left) and in 1-dimension for $\beta_{s}^{J/\psi\phi}$ (right). The SM predicted value is marked with a black point or line.}
\label{fig:results1}
\end{figure}

With the current data sample it is not possible to quote a point value for the phase $\beta_{s}$ due to the symmetries in the likelihood function and the non-Gaussian error distribution. Instead the results are presented as frequentist likelihood contours; a profile-likelihood ratio ordering technique is used to ensure full coverage. Figure~\ref{fig:results1} shows the likelihood contours in the $\beta_{s} - \Delta\Gamma$ plane. 
%
The confidence interval for $\beta_s^{J/\psi\phi}$ at the 68 \% confidence level is $[0.02,0.52] \cup [1.08,1.55]$ and at the 95 \% confidence level, $[-\pi/2, -1.44] \cup [-0.13,0.58] \cup [0.89,\pi/2]$.

The upper limit on the fraction of the $S$-wave $KK$ ($f_0$) component in the $B_s^0 \rightarrow J/\psi\phi$ signal was measured to be $<$ 6.7\% at the 95\% confidence level.  
%

In the hypothesis of no $CP$ violation ($\beta_s =0.0$), the values of the $B_{s}^{0}$ lifetime, $\tau_{s} = 1.53 \pm 0.025~\rm{(stat.)}\pm0.01~\rm{(syst.)}~\rm{ps}$, decay width difference $\Delta\Gamma_{s} = 0.075 \pm 0.035~\rm{(stat.)}\pm 0.01~\rm{(syst.)}~\rm{ps}^{-1}$ the transversity amplitudes, $|A_{\parallel}|^{2} = 0.231 \pm 0.014~\rm{(stat.)}\pm 0.015~\rm{(syst.)}$ and $|A_{0}|^{2} = 0.524 \pm 0.013~\rm{(stat.)}\pm 0.015~\rm{(syst.)}$ and the strong phase $\phi_{\perp} = 2.95  \pm 0.64~\rm{(stat.)}\pm 0.07~\rm{(syst.)}$  from the flavour tagged fit are determined~\cite{CDFnew}.

%

\section{Conclusions}

Previous measurements from both the CDF~\cite{CDFpub} and D\O~\cite{D0pub} experiments, and a combined Tevatron result~\cite{comb}, indicated a shift of about 2 $\sigma$ from the Standard Model expectation for the $B_s^0$ mixing phase.  The new $\beta_s^{J/\psi\phi}$ measurement from the CDF experiment shows good agreement with the Standard Model expected value, with a $p$-value of 44\%, equivalent to a deviation of 0.8 $\sigma$. 
These results benefit from an improved selection and particle ID to take advantage of the full available statistics, updated flavour tagging, and the inclusion of the $S$-wave $KK$ component of the signal fraction in the fit. 
Additionally, results include the world's most precise single measurements of the $B_s^0$ lifetime and decay width difference in the hypothesis of no $CP$ violation. 

Future prospects for this measurement at CDF include doubling again the data sample, the potential inclusion of data from different trigger paths and the addition of other $B_s^0$ decay channels to the analysis as well as continuous improvements to the analysis technique.

\section{Acknowledgements}
The author would like to thank the University of Pittsburgh for providing funding to attend at FPCP 2010.

\end{document}